\documentclass[modern]{aastex62}

\graphicspath{{./}{figures/}}
\usepackage{amssymb}
\usepackage{amsmath}
\usepackage{float}
\received{6 May 2021}
\revised{12 July 2021}
\accepted{12 July 2021}
\submitjournal{\apj}

\shorttitle{Interferometric Observations of HD 97658}
\shortauthors{Ellis et al.}

\makeatother
\newcommand{\thisstar}{HD~97658 }
\newcommand{\thisplanet}{HD~97658~b }
\newcommand{\udd}{\ensuremath{0.296\pm0.004}}
\newcommand{\ldd}{\ensuremath{0.314\pm0.004}}
\newcommand{\rstar}{\ensuremath{0.728\pm0.008}}
\newcommand{\lstar}{\ensuremath{0.351\pm0.007}}
\newcommand{\fbol}{\ensuremath{2.42\pm0.05\times10^{-8}}}
\newcommand{\temp}{\ensuremath{5212\pm43}}
\newcommand{\mulimb}{\ensuremath{0.629\pm0.014}}
\newcommand{\Ld}{\ensuremath{\mathcal{L}}}
\newcommand{\mstardir}{\ensuremath{0.85\pm0.08}}
\newcommand{\rhostardir}{\ensuremath{3.1\pm0.3}}

\newcommand{\unid}{$\theta_\text{UD}$ }
\newcommand{\angd}{$\theta_\text{LD}$ }
\providecommand{\msun}{\ensuremath{\,\text{M}_\Sun}}
\providecommand{\mearth}{\ensuremath{\,\text{M}_\Earth}}
\providecommand{\rsun}{\ensuremath{\,\text{R}_\Sun}}
\providecommand{\rearth}{\ensuremath{\,\text{R}_\Earth}}
\providecommand{\lsun}{\ensuremath{\,\text{L}_\Sun}}
\providecommand{\teff}{$T$\textsubscript{eff}}
\providecommand{\funits}{\,erg~s\textsuperscript{-1}~cm\textsuperscript{-2}\,}

\begin{document}

\title{Directly Determined Properties of HD 97658 from Interferometric Observations}
\correspondingauthor{Tyler Ellis}
\email{telli17@lsu.edu}

\author[0000-0001-6584-9919]{Tyler G. Ellis}
\affil{Louisiana State University \\
202 Nicholson Hall \\
Baton Rouge, LA 70803}

\author[0000-0001-9879-9313]{Tabetha Boyajian}
\affil{Louisiana State University \\
202 Nicholson Hall \\
Baton Rouge, LA 70803}

\author[0000-0002-5823-4630]{Kaspar von Braun}
\affiliation{Lowell Observatory\\
1400 W. Mars Hill Road\\
Flagstaff, AZ 86001, USA}

\author[0000-0002-6257-6051]{Roxanne Ligi}
\affiliation{INAF-Osservatorio Astronomico di Brera \\
Via E. Bianchi 46 \\
I-23807 Merate, Italy}

\author[0000-0001-7425-5055]{Denis Mourard}
\affiliation{Universit\'e C\^ote d'Azur, Observatoire de la C\^ote d'Azur, CNRS, Laboratoire Lagrange, Parc Valrose, 06108 Nice Cedex, France}

\author[0000-0003-2313-467X]{Diana Dragomir}
\affiliation{Department of Physics and Astronomy, University of New Mexico, 1919 Lomas Blvd NE, Albuquerque, NM 87131, USA}

\author[0000-0001-5415-9189]{Gail H. Schaefer}
\affiliation{The CHARA Array of Georgia State University, Mount Wilson, CA 91023, USA}

\author[0000-0001-9939-2830]{Christopher D. Farrington}
\affiliation{The CHARA Array of Georgia State University, Mount Wilson, CA 91023, USA}

\begin{abstract}
We conducted interferometric observations with the CHARA Array of transiting super-Earth host HD 97658 and measured its limb-darkened angular diameter to be $\theta_{\text{LD}}=0.314\pm0.004$~mas. The combination of the angular diameter with the Gaia EDR3 parallax value with zero-point correction ($\pi=46.412\pm0.022$~mas, $d=21.546\pm0.011$~pc) yields a physical radius of $R_\star=0.728\pm0.008$~\rsun. We also measured the bolometric flux of the star to be $F_\text{bol}=2.42\pm 0.05\times 10^{-8}$~erg~s\textsuperscript{-1}~cm\textsuperscript{-2}, which, together with angular size, allows a measurement of the effective temperature $T_{\text{eff}}$~=~$5212\pm43$~K. Our directly determined physical stellar properties are in good agreement with previous estimates derived from spectroscopy. We used our measurements in combination with stellar evolutionary models and properties of the transit of HD 97658 b to determine the mass and age of HD 97658 as well as constrain the properties of the planet. Our results and our analysis of the TESS lightcurve on the planet (TOI-1821) corroborate previous studies of this system with tighter uncertainties.
\end{abstract}


%%%%%%%%%%%%%%%%%%%%%%%%%%%%%%%%%%%%%%%%%%%%%%%%%%%%%%%%%%%%%%%%%%%%%%%%%%%%%%%%
\section{Introduction} \label{sec:intro}
%%%%%%%%%%%%%%%%%%%%%%%%%%%%%%%%%%%%%%%%%%%%%%%%%%%%%%%%%%%%%%%%%%%%%%%%%%%%%%%%
Interferometric observations of stars provide a unique opportunity to directly measure one of the most fundamental parameters of the star: its radius. Interferometry achieves the resolution of an extremely large telescope by combining the light from one or multiple pairs of telescopes across a variety of separations, or baselines. In particular, optical/near-infrared interferometry requires baselines of only tens of meters to achieve resolutions of milli-arcseconds (mas). Direct measurements of stellar radii at great precision will in turn reduce uncertainty in other derived stellar parameters (e.g. effective temperature, surface gravity, or density). Direct observations of stellar radii have highlighted a systematic discrepancy between evolutionary models and reality. \citet{Boyajian2012} has shown that stellar evolutionary models underestimate radii by $\sim 5\%$ and overestimate temperatures by $\sim 3\%$ for K and M dwarfs.

Observations with interferometric arrays play an important role in understanding as well as refining exoplanet system properties. In order to understand the properties of the exoplanet, the properties of the star must first be well constrained. In particular, transiting exoplanets provide a measure of the planet's radius, but this measurement is in units of the host star's radius. Any uncertainty or bias in stellar radius will propagate into estimates of the planet's equilibrium temperature, density, habitability, and composition. Interferometry gives a direct measurement of the stellar radius with little or no dependence on stellar models. This technique has been used in the literature to refine the properties of several important systems, such as 55 Cancri which hosts five radial velocity exoplanets including another transiting super-Earth \citep{vonBraun11} and transiting exoplanet host star GJ 436 where evolutionary models underestimated the stellar radius by $\sim11\%$ \citep{vonbraun2012}. There have also been multiple interferometric surveys of large numbers of exoplanet host stars such as \citet{Baines2008}. The field of interferometry has also seen incredible developments in the field of imaging and astrometry with the ESO GRAVITY project \citep{GRAVITY2017}. In 2019, the GRAVITY collaboration announced the first spectrum of an exoplanet observed with interferometry and refined the astrometric position with 100$\mu$as precision \citep{GRAVITY2019}.

The exoplanet host star \thisstar is of particular interest in this regard as it is the home of a transiting super-Earth \citep{dragomir2103, Howard2011}. \thisstar is a bright, $m_V$=7.78~mag K1 dwarf with a moderately low iron content of [Fe/H]$=-0.23$~dex, which was discovered to have a Neptunian mass exoplanet by the NASA Eta-Earth Keck-HIRES radial velocity survey \citep{Howard2011}. Follow-up time series observations with the \textit{Spitzer} and \textit{MOST} space telescopes detected a transit whose depth indicated an estimated planetary radius of a few Earth radii \citep{dragomir2103,vangrootel2014}. These properties together make \thisplanet a so-called super-Earth (planets with radii of 1--4~\rearth\ and masses of 1--10~\mearth; \citealt{Bryan2019}). Super-Earths can take the form of water worlds with a smothering dense atmosphere or rocky behemoths with minimal atmospheres, both often consistent within the uncertainties of planetary mass and radius \citep{dragomir2103}. Super-Earths captivate planetary scientists as they are the most populous of observed exoplanets (30--50\% of Sun-like stars host one or more super-Earths \citep{Bryan2019}), however they do not exist within our own solar system and must be studied solely as exoplanets.

We are interested in refining the properties of \thisplanet by directly measuring the host star's properties. In \textsection \ref{sec:obs}, we describe the interferometric observations of HD 97658. In \textsection \ref{sec:properties}, we report the resulting directly measured angular diameter, bolometric flux. In \textsection \ref{sec:derived}, we model and measure the mass of the star. We then derive updated properties of \thisstar and its planet using the TESS lightcurve and our measured results. Lastly, in \textsection \ref{sec:conclusion} we summarize and conclude this work.

%%%%%%%%%%%%%%%%%%%%%%%%%%%%%%%%%%%%%%%%%%%%%%%%%%%%%%%%%%%%%%%%%%%%%%%%%%%%%%%%
\section{Interferometric Observations} \label{sec:obs}
%%%%%%%%%%%%%%%%%%%%%%%%%%%%%%%%%%%%%%%%%%%%%%%%%%%%%%%%%%%%%%%%%%%%%%%%%%%%%%%%
We observed \thisstar over the course of several nights using the Georgia State University Center for High Angular Resolution Astronomy (CHARA) Array at the Mount Wilson Observatory using the Classic (near-IR), VEGA (optical), and PAVO (optical) beam combiners \citep{theo2005,Mourard2009,Ireland2008}.  A summary of the observations is found in Table~\ref{tab:obs}.

An interferometer measures visibilities ($V$), which quantifies the contrast of the dark and bright parts of the interference fringe pattern. In practice, this is the contrast of the time-averaged minimum and maximum power of the fringe pattern (see \citet{Lawson2000} for a full description of what an interferometer measures in Chapter 2.6).

As the visibilities measured at the time of observing include instrumental and atmospheric  effects, it is necessary to observe stars with predictable visibilities to calibrate the data. These calibrator stars are observed in sequence with a science star and allow a measure of the combined systematic effects. Calibrators stars are of a known, and ideally unresolved, size.

Each observation consists of bracketed sequences of the form calibrator 1 - science target - calibrator 2 (or calibrator 1 again), and then reverse. One bracket is one observation of the target sandwiched by the calibrators.  

As the beam combiners used in this work observe in different bandpasses, it is sometimes necessary to use different calibrators for each instrument. Wherever possible, we calibrated the raw square visibilities of the calibrators against each other \citep{vonBraun2017}. This provides insight into previously unknown duplicity, activity, or other anomalous behaviors in a calibrator.  No stars were expunged from our list of acceptable calibrators for any reason. The limb darkened angular size of the calibrators and magnitudes are summarized in Table~\ref{tab:cals}.

We chose calibrators for this work using the JMMC Stellar Diameter Catalog (JSDC) version 2 \citep{jmmc2016,duvert2016}\footnote{\url{http://www.jmmc.fr/catalogue\_jsdc.htm}}.  Ideally, we restrict our search for calibrators that are not resolved on the baselines used with beam combiner's bandpass. Unresolved sources have predicted squared visibilities $V^2\gtrapprox0.9$. Further, we discard potential calibrators that have known duplicity and/or have rapid rotation driving equatorial distention. Lastly, preference is given to calibrators of comparable brightness and with a minimal angular separation in the sky from the science target.

Calibrator selection is a nontrivial process which can affect the uncertainties in final calibrated data. \citet{vanBelle2005} demonstrated that a 5\% uncertainty in the calibrator diameter can propagate up to a $\sigma_{V^2}\sim0.04$ uncertainty in the calibrated visibilities, though the amount depends on interferometer configuration, instrument, and calibrator size. Oftentimes, for small, faint targets such as HD 97658, most calibrators which are sufficiently smaller so as to be unresolved are often too faint to be observed. As such, it is sometimes necessary to compromise for partially resolved calibrators. The caveat with such a compromise is that the target's diameter will only be known as well as the calibrators.  In this work, the calibrators HD 101688 and HD 96738 are slightly resolved using the PAVO/VEGA beam combiner and S1-W1 baseline, $V^2\simeq 0.6$.  Thus, the predicted angular size of these calibrators are of concern to the error budget. However, within this size regime, $\theta<0.45$~mas, the dominant source of error is in the measurement of the visibilities rather than assumptions of calibrator size \citep{vanBelle2005}. 

We used the \textit{isoclassify}\footnote{\url{https://github.com/danxhuber/isoclassify}} program to estimate the angular diameters of these calibrators independently of the JMMC's surface brightness relationship method \citep{Huber2017isoclassify1,Huber2020isoclassify2}. We find the \textit{isoclassify} and JMMC diameters to be consistent, though the adopted \textit{isoclassify} diameters had larger, more conservative uncertainties (an average of 1.9$\times$ larger). All of differences in size are less than 1$\sigma$, with the average statistical tension $Avg\left( (|\theta_\text{iso}-\theta_\text{JMMC}|)/\sqrt{\sigma_\text{iso}^2+\sigma_\text{JMMC}^2}\right)=0.33$.

As previously mentioned, the greater uncertainty in the calibrated visibilities from using resolved calibrators is propagated forward into the uncertainty of the fitted angular diameter. We also note that data taken with the VEGA on several nights are calibrated with only a single star. However, our analysis finds consistency in the calibrated results from night to night and between PAVO and VEGA, assuring confidence in our choice of calibrators and calibration methods \citep{Ligi2016, Baines2018, Lachaume2019}. 

In order to complete an observation, the interferometric fringes must be found by equalizing the optical path length from the stars to the beam combiner through the two telescopes. The fringes are found by scanning the additional path length up and down for one of the telescopes until fringes are detected. In good conditions, scanning initially takes about 5--15 minutes per star and about twice that in difficult observational conditions. Finding the fringe packet for subsequent observations of the same star typically goes much faster, at most a few minutes.

Observations with the Classic beam combiner consist of approximately 2.5 minutes of integration in the \textit{H} band. Shutter sequences proceed and follow integration. Classic data are reduced using the \texttt{REDFLUOR} package to produce raw squared visibilities which are then calibrated using the \texttt{CALIBIR} package --- both of these software routines are provided as binary executables from CHARA \footnote{\url{http://www.astro.gsu.edu/~theo/chara_reduction/index.html}}. Classic observes in a single spectral channel at a time and gathers a single data point per bracket.

Observations with the PAVO beam combiner consist of around 2 minutes integration followed by approximately 3 minutes of shutter sequences and dark integration for reduction and calibration. PAVO  data are reduced and calibrated using IDL routines which are also provided by CHARA \citep{Ireland2008}\footnote{\url{https://gitlab.chara.gsu.edu/fabien/pavo.git}}. As PAVO is an integral field unit, it collects a spectra of fringes with each bracket between 630-950 nm bandwidth (resolution R = 30) \citep{Ireland2008}. 

Observations with VEGA require more time. Calibrators are observed during $\sim$15 min while the science star is generally observed during $\sim$30 min to ensure enough signal. The data are then reduced using the \texttt{vegadrs} pipeline \citep{Mourard2009, Mourard2011}. VEGA's bandpass is broken into 20~nm bins and creates two data points per baseline at a time. 

\begin{table}[ht]
  \begin{center}
  \begin{tabular}{ccccc}
  Date [UT]		&	Baseline	&	Combiner 	&	Brackets	&	Calibrators\\\hline\hline
  2015-02-04	&	S1/W1		&	Classic		&	5			&	HD 93152, HD 99267\\
  2015-02-05	&	S1/W1		&	Classic		&	7			&	HD 95804, HD 99267\\
  2015-02-11	&	S1/W1		&	PAVO		&	4			&	HD 101688, HD 96738\\
  2017-03-13    &   W1/W2       &   VEGA        &   2           &   HD 89239\\
  2017-03-17    &   S2/W2       &   VEGA        &   2           &   HD 107168\\
  2018-04-28    &   E2/W2       &   VEGA        &   2           &   HD 97638\\
  2019-05-05    &   S1/W2       &   VEGA        &   3           &   HD 96738\\
  2020-03-05    &   S2/E2       &   VEGA        &   4           &   HD 96738, HD 107168
  \end{tabular}
  \caption{Summary of observations of \thisstar. Both PAVO and Classic instruments were used in the single baseline mode of the CHARA array, while VEGA uses multiple baselines simultaneously. See \citet[Table 1]{theo2005} for a complete description of the available baselines. Each bracket corresponds to one observation of the target. See \S~\ref{sec:obs} for further details \label{tab:obs} }
  \end{center}
\end{table}

\begin{table}[ht]
    \centering
    \begin{tabular}{cccc}
         Calibrator & \textit{V} mag & \textit{H} mag & \angd [mas] \\\hline\hline
         HD 89239   &   6.530        &   6.599	      &   0.159$\pm$0.008\\
         HD 93152   &   5.285        &   5.384        &   0.279$\pm$0.012\\
         HD 95804   &   6.766        &   6.288        &   0.208$\pm$0.005\\
         HD 96738   &   5.593        &   5.442        &   0.269$\pm$0.033\\
         HD 99267   &   6.606        &   6.091        &   0.241$\pm$0.006\\
         HD 101688  &   6.291        &   5.730        &   0.281$\pm$0.011\\
         HD 107168  &   6.220        &   5.969        &   0.241$\pm$0.011
    \end{tabular}
    \caption{Summary of calibrator stars used in all observing campaigns. All photometry data are taken from the JMMC Stellar Diameters Catalog v2. The cited angular diameters are limb darkened and estimated using \textit{isoclassify}.}
    \label{tab:cals}
\end{table}

%%%%%%%%%%%%%%%%%%%%%%%%%%%%%%%%%%%%%%%%%%%%%%%%%%%%%%%%%%%%%%%%%%%%%%%%%%%%%%%%
\section{Directly Determined Stellar Properties} \label{sec:properties}
%%%%%%%%%%%%%%%%%%%%%%%%%%%%%%%%%%%%%%%%%%%%%%%%%%%%%%%%%%%%%%%%%%%%%%%%%%%%%%%%

  \subsection{Stellar Diameter} \label{sec:diam}
  
  The calibrated squared visibilities $V^2$ can be fit to the radial profile of the 2D Fourier pair of a uniformly illuminated disk or a limb darkened disk, $\theta_{\text{UD}}$ and $\theta_{\text{LD}}$, respectively. The resulting profile is a function of the projected baseline \textit{B}, observational wavelength $\lambda$, and most importantly the angular size of the object \citep{Hanbury1974}. The functional form, shown in Eq.~\ref{eq:vis2} is a combination of Bessel functions of the first kind, $J_\alpha(x)$, scaled with the linear limb darkening coefficient $\mu$.
  \begin{equation}
  	\label{eq:vis2} V^2 = \left [ \left ( \dfrac{1-\mu}{2}+\dfrac{\mu}{3}\right )^{-1} \cdot \left ( (1-\mu)\cdot\dfrac{J_1(x)}{x}+\mu \sqrt{\dfrac{\pi}{2}}\cdot\dfrac{J_{3/2}(x)}{x^{3/2}}\right )\right ]^2
  \end{equation}
  
  \noindent where $x=\dfrac{\pi B \theta}{\lambda}$.
  
\begin{figure}[ht]
  	\begin{center}
  		\plotone{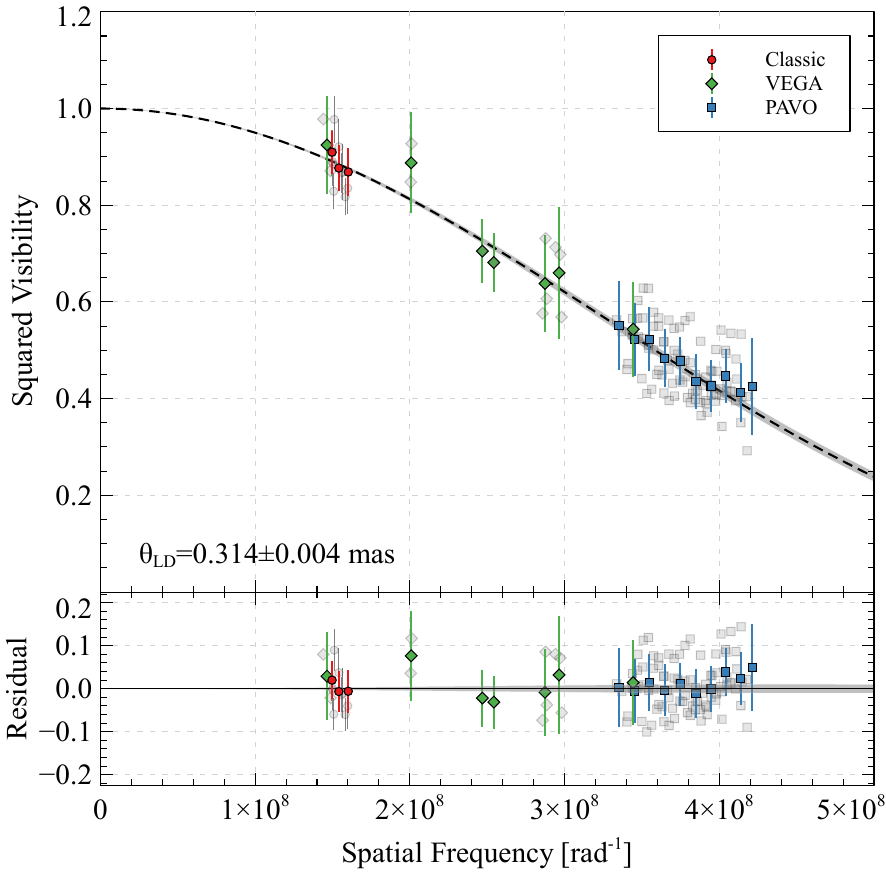}\caption{\label{fig:visibilities}Top panel shows the calibrated squared visibilities and uncertainties from CHARA observations of \thisstar\!. We binned and averaged the data by equal spacing in of 10$^7$~rad\textsuperscript{-1} for clarity, and plot these points over all of the calibrated data shown in transparent grey. Classic observations are shown as red circles grouped on the left at lower spatial frequencies, VEGA observations are green diamonds in the middle, and PAVO observations are blue squares grouped on the right at higher spatial frequencies. The fitted visibility curve for a limb darkened disk is shown as a dashed line with the parameters found in \textsection \ref{sec:diam}. We show the uncertainty in the angular diameter as the grey region around the best fit curve. The residuals for the fit are shown in the bottom panel. See \S~\ref{sec:diam} for details.}
  	\end{center}
  \end{figure}

    We fit the uniform disk model, Eq. \ref{eq:vis2} with $\mu=0$, for the combined datasets using the \textit{Scipy curve\_fit} nonlinear least-squares minimization routine \citep{Jones2001}. As part of the fitting process, we find the optimized fit for many realizations of the dataset by sampling the wavelength solution uncertainty. The resulting distribution of fits for the uniform disk is \unid~=~\udd~mas. We performed the fitting routine for the VEGA and PAVO datasets separately as a check for consistency and found the best fit uniform disk angular diameters of $\theta_{\rm UD,VEGA}$~=~$0.282\pm0.024$~mas and $\theta_{\rm UD,PAVO}~=~0.296~\pm~0.004$~mas. As seen in Fig. \ref{fig:visibilities}, there is more internal spread amongst the VEGA data than the Classic or PAVO data, which drives the uncertainty up slightly. The enlarged uncertainties are likely caused by dividing the starlight of this faint target amongst the multiple simultaneous baselines of VEGA. We opt not to perform this same fit with the Classic data as \thisstar is not well resolved by the instrument, but can still act as a sanity check for the other instruments. The uniform disk angular diameter and other stellar properties are summarized in Table~\ref{tab:props}.
  
  We fit the limb darkened disk model using the same technique as above, adding in Monte Carlo realizations of the limb darkening parameter $\mu$ as part of the fitting process. As all visibility curves are near unity at low spatial frequencies, it is safe to combine the Classic data with the VEGA/PAVO data. We estimated the limb darkening parameter $\mu$ using the Limb Darkening Toolkit (LDTK) \citep{LDTK1,LDTK2}. Throughout this work we find the limb darkening coefficient and associated limb darkened diameter in the Bessel R filter. We provide the LDTK module with estimates of \teff, $\log(g)=4.5\pm0.1$, $Z=0.03\pm0.01$. We iterate the fitting process twice to reflect our refined measurements in \teff\, which goes into the estimation of the limb darkening coefficient. The first run uses the fit of \unid with our measurement of the bolometric flux to estimate \teff, as is discussed in the following section \S~\ref{sec:fbol_teff}. Then we use the results from the first fit of \angd to estimate \teff\, and find our new estimation of $\mu$. With the final estimation of $\mu$, we then find our final fit of \angd. We scale the uncertainties in $\mu$ by a factor of 5 during each iteration of the fitting process as the LDTK distribution seems unrealistically tight when compared to \citet{Claret2011}, though this does not significantly contribute to the final error budget. Our final estimation of the linear limb darkening coefficient in the Bessel R filter is $\mu_\text{R}=\mulimb$.
  
  The iteration process resulted in a best fit limb darkened angular diameter of \angd~=~\ldd~mas. The reduced chi-square for the linear limb darkened model is $\chi^2_{\nu=113}=0.934$ ($p=0.68$), indicating good agreement between theory and observation.  This fit and the calibrated squared visibilities are shown in Figure~\ref{fig:visibilities} with the uncertainties scaled to fix $\chi^2_\nu$ at unity.

  We calculate the physical radius of \thisstar as \rstar~\rsun\, using the Gaia EDR3, zero-point corrected parallax measurement (Table~\ref{tab:props}). Applying the zero-point correction, the Gaia parallax of 46.412$\pm$0.022~mas yields a corresponding distance of $d=21.546\pm0.011$~pc \citep{GAIAEDR3,Lindegren2021}. Previously published radius estimates include \citet{dragomir2103} who obtained an estimate of $0.703^{+0.035}_{-0.030}$~\rsun\ from evolutionary models fit within \texttt{EXOFAST} (\citet{eastman2013}) and \citet{vangrootel2014} who derived $0.741^{+0.024}_{-0.023}$~\rsun\ from the spectroscopic temperature, bolometric correction, and Hipparcos parallax. It is of interest that the estimation of the radius using the evolutionary models in \texttt{EXOFAST}  is lower than the spectroscopic radius and the directly measured radius in this paper. As has been explored in \citet{Boyajian2012}, evolutionary models systematically underestimate the radius by a few percent. The high resolution capabilities of the interferometer complemented with the exquisite parallax measurements from Gaia allow us to report a physical radius with about 1/3 the uncertainty of prior works. The uncertainty in our physical radius predominantly comes from the fit of the angular diameter, with the parallax contribution essentially negligible.

  \subsection{Bolometric Flux and Temperature} \label{sec:fbol_teff}
  We fit an interpolated K0.5 Pickles \citep{Pickles1998} template spectrum to collected literature measurements of broadband photometry to measure the bolometric flux. Photometry used in this fit include measurements in Johnson \textit{UBV}, Cousins $R_c, I_C$, 2Mass \textit{JHK}.  The photometric measurements are from \citet{vanLeeuwen2007, Skrutskie2006, Kotoneva2002, Koen2010, Bessell2000, Kharchenko2001, Droege2006, Mermilliod1994}.
  
  Interstellar extinction was fixed at 0 in the fitting routine as the proximity of the star should render any extinction effects negligible. The fit was also performed with $A_V$ as a free parameter. This modification found a value of $A_V=0.027\pm0.015$, so we accept our original assumption of no extinction for this work.
  
\begin{figure}[H]
      \centering
      \plotone{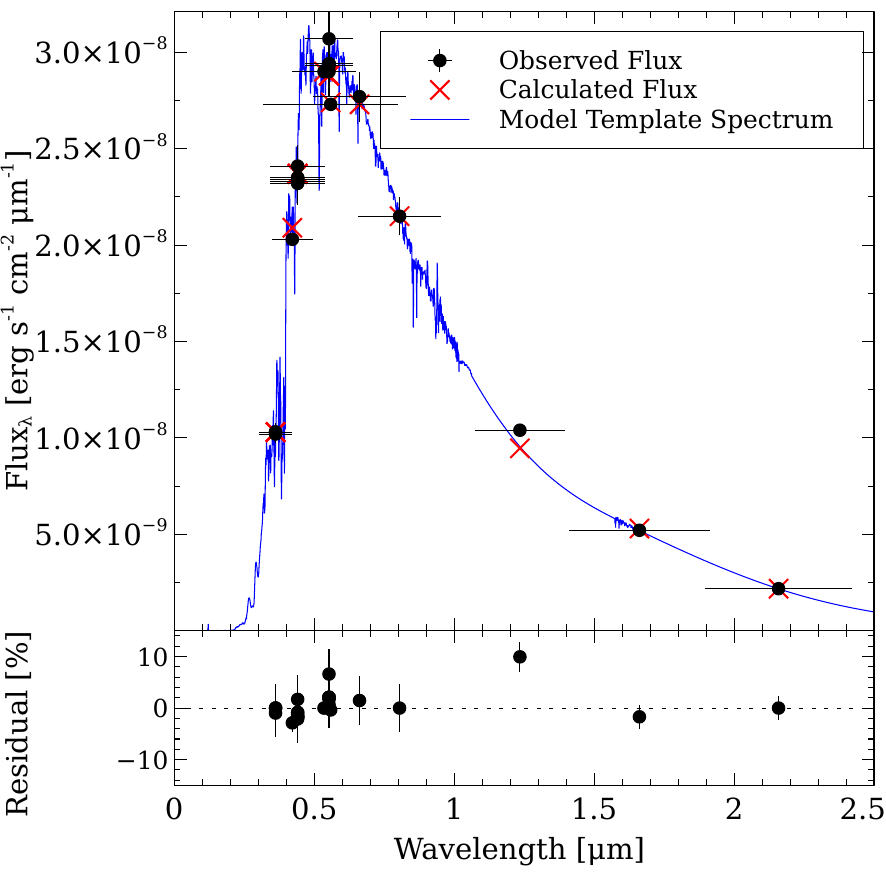}
      \caption{The template spectrum shown in blue is an interpolated Pickles K0.5V stellar spectrum with nominal metallicity. We used literature broadband photometry and associated uncertainties and bandpass widths to scale the spectrum. The photometry is shown as the black points. The horizontal bars on the photometry data points represent the width of the filter bandpass. This scaled spectrum was then integrated to yield the bolometric flux. We calculate the scaled model flux values at the center of the bandpasses and show them as red crosses here. The bottom panel shows the percent residual of this fit. See \S~\ref{sec:fbol_teff} for discussion.}
      \label{fig:SED}
  \end{figure}

  The template spectra was then scaled to fit to the photometric measurements and integrated to obtain a bolometric flux $F_{\text{bol}}$=\fbol~\funits. We use the updated filter profiles and zero-point corrections as discussed in \citet{mann2015filters}. Further, we account for unknown systematics by applying a 2\% addition in quadrature to the bolometric flux uncertainty as is suggested in \citet{Bohlin2014}. Our flux measurement and the parallax distance gives a stellar luminosity of $L$=\lstar~\lsun. The assembled photometry and spectral fit are shown in Figure~\ref{fig:SED}. This model has a goodness of fit of $\chi^2_{\nu=29}=0.79$ ($p=0.78$). Further details of this technique are available in \citet{vanBelle2007,vonBraun2017}. We extended the infrared portion of the Spectral Energy Distribution (SED) ($\lambda>12.5$~\micron) with the WISE W 1--4 data to check for an infrared excess, but did not observe any such excess \citep{Wright2010}. We note that \thisstar\ is saturated in W1 and W2, so we opted to exclude all of these data from the SED fit.
  
  The Stefan-Boltzmann equation can be rewritten to express the temperature in terms of the observables --- bolometric flux and angular diameter:
  \begin{equation}\label{eq:SB}
      T_\text{eff} = 2341 \cdot \left( \dfrac{F_{\rm bol}}{\theta_{\rm{ LD}}^2}\right)^{1/4} \text{ K},
  \end{equation}
  where the bolometric flux is in units of $10^{-8}$~\funits and the angular diameter is in mas. We determined an effective temperature of \teff=\temp~K given the limb darkened diameter and above bolometric flux. \citet{Howard2011} measured the effective temperature of \thisstar as $5170\pm44$~K which was determined from spectroscopy. This temperature is lower than what we found, but consistent with our result at the 1$\sigma$ level.

%%%%%%%%%%%%%%%%%%%%%%%%%%%%%%%%%%%%%%%%%%%%%%%%%%%%%%%%%%%%%%%%%%%%%%%%%%%%%%%%
\section{Derived Stellar and Planetary Properties} \label{sec:derived}
%%%%%%%%%%%%%%%%%%%%%%%%%%%%%%%%%%%%%%%%%%%%%%%%%%%%%%%%%%%%%%%%%%%%%%%%%%%%%%%%
Using the measured properties above, we compute here other properties of the star and its planet.  As the host star plays an important role in determining many of the properties of the planet, we can also update several of the estimated properties of the planet with our improved stellar parameters. We also include new observations of the planet's transit with TESS. A summary of all the following results is shown in Table~\ref{tab:props}.

\subsection{Age and Mass Estimation with Isochrones} \label{sec:ageandmass}
\begin{figure}[ht]
	\begin{center}
		\plottwo{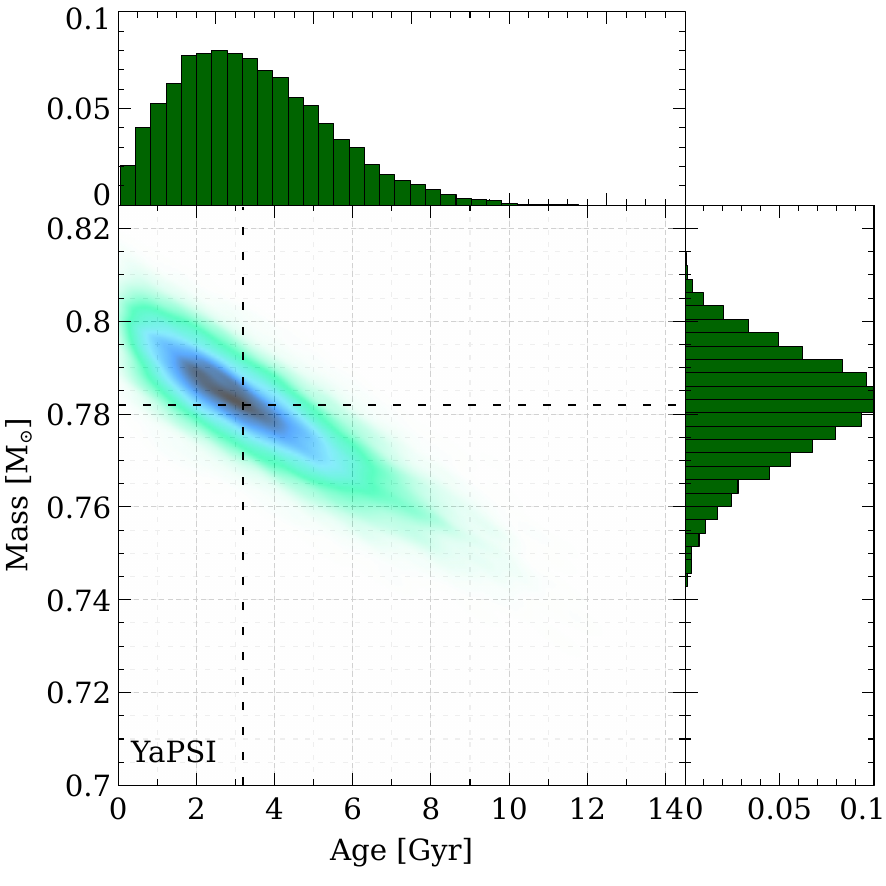}{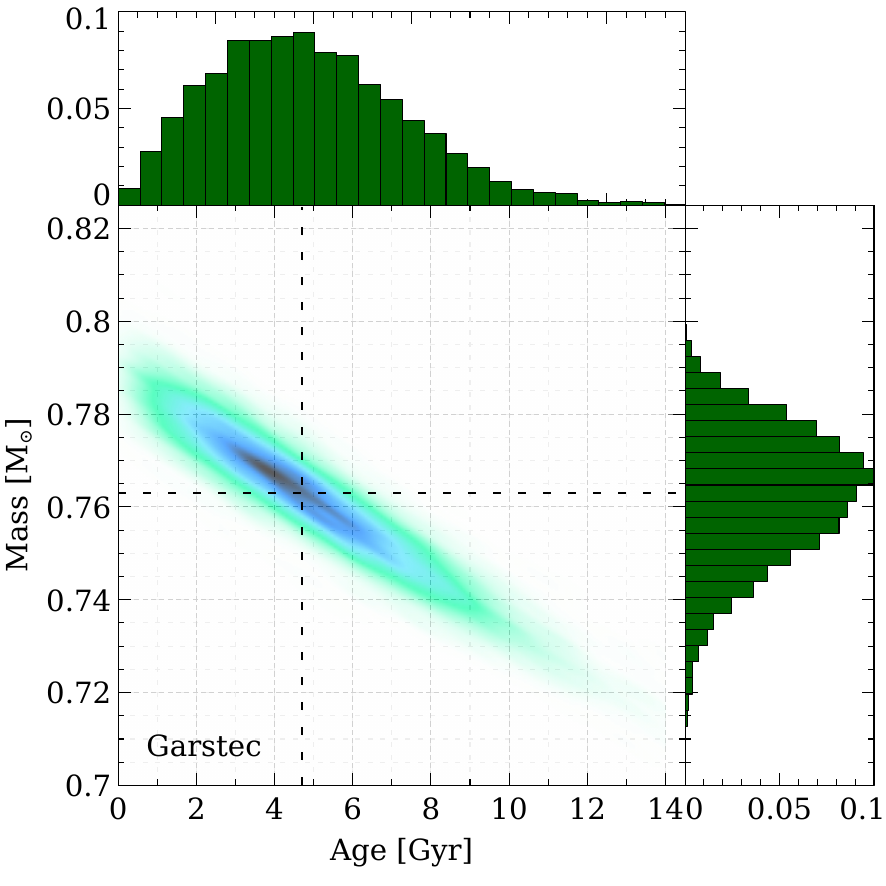}\caption{\label{fig:yapsi}Posterior distribution of age and mass from the \textit{bagemass} Bayesian evolutionary track fitting program as discussed in \S~\ref{sec:derived}. The YaPSI model output is shown on the left and the Garstec model on right. Higher posterior density is shown as the darker regions in the bottom left plot. The posterior population distribution for the ages is shown on the top and the posterior population distribution for the mass is shown on the right. Black dashed vertical/horizontal lines indicate the median for the age ($3.2\pm1.9$~Gyr, $4.7\pm2.5$~Gyr) and mass ($0.782\pm0.011$~\msun, $0.763\pm0.011$~\msun) for the YaPSI and Garstec models, respectively. We combine the solutions to derive a final estimated age $3.9_{-2.03}^{+2.6}$~Gyr and mass  $0.773_{-0.018}^{+0.015}$~\msun\ as described in \S~\ref{sec:derived}. }
	\end{center} 
\end{figure}

We use two stellar evolutionary models to estimate the stellar age and mass. The Garstec and YaPSI stellar evolutionary models \citep{Spada2017, Weiss2008} were fit using the \texttt{bagemass} Fortran program \citep{Maxted2015}. These models use our directly measured temperature, the luminosity inferred from the parallax and bolometric flux, the spectroscopically determined [Fe/H] from \citet{Howard2011}, and the density inferred from the transit discussed later in this section as predictors of goodness of fit. We assumed uniform priors on age, mass, and surface [Fe/H].  One of the issues with estimating age using isochrones for this range of stellar masses is a lower mass star afforded a longer time to evolve can have the same observable properties as a higher mass star that is younger. This bias is reflected as the elongated distribution in age and mass. We show the posterior distribution for the YaPSI and Garstec models in Figure~\ref{fig:yapsi}.
   
We then checked the consistency of the two model's predicted age and mass using a $\chi^2$-test. We first combined (i.e. summed) the two posterior distributions for the ages/masses and computed the resultant median age and mass. We then compute:
\begin{equation}
    \chi^2=\sum_{i=1}^2 \frac{x_i-\bar{x}^2}{\sigma_i^2}
\end{equation}
where $x_i$ is the median mass/age from the two models, $\bar{x}$ is the median from the combined distribution, and $\sigma$ is their associated uncertainties. Assuming that these are drawn from a $\chi^2$ distribution, the confidence in the agreement of the two posterior distributions can be computed with:
\begin{equation}
        p=\int_{\chi^2}^{\infty} f(\chi^2) d\chi^2
\end{equation}

The $\chi^2$ distribution, $f(\chi^2)$, for this case has $2-1=1$ degrees of freedom, for the two samples of the age/mass. The computed p-value for the ages is p=0.62 and p=0.27 for the masses. Both the mass and age distributions agree within 95\% confidence, so we conclude that the median and $1\sigma$ confidence interval of the combined distribution is representative of both. The combined isochrone models estimated the age of \thisstar as $3.9_{-2.03}^{+2.6}$~Gyr and the mass as  $0.773_{-0.018}^{+0.015}$~\msun.
   
\thisstar has a chromospheric Ca II H and K activity index of $\log\left(R^\prime_{HK}\right)=-4.971$ \citep{Isaacson2010}. \citet{Isaacson2010} then used the relationship from \citet[Eq. 3]{Mamajek2008} to estimate an age of $6.06\pm0.91$~Gyr\footnote{In order to estimate uncertainty on this age, we applied the suggested RMS of 0.07 dex on $\log\tau$ from \citet{Mamajek2008}}. Using the Gaia $G_{BP} - G_{RP}=0.843$ and the rotation period of 34$\pm$2~days from \citet{guo2020}, we find another estimate of the age of \thisstar of $6.25\pm0.56$~Gyr using gyrochronology with \texttt{stardate} \citep{Angus2019}. Both of these techniques agree within $\sim1\sigma$ of the combined isochrone model age from this work.

\subsection{Exoplanet Modeling with \textit{TESS}} \label{sec:exoplanet}
\begin{figure}[ht]
    	\begin{center}
    		\plotone{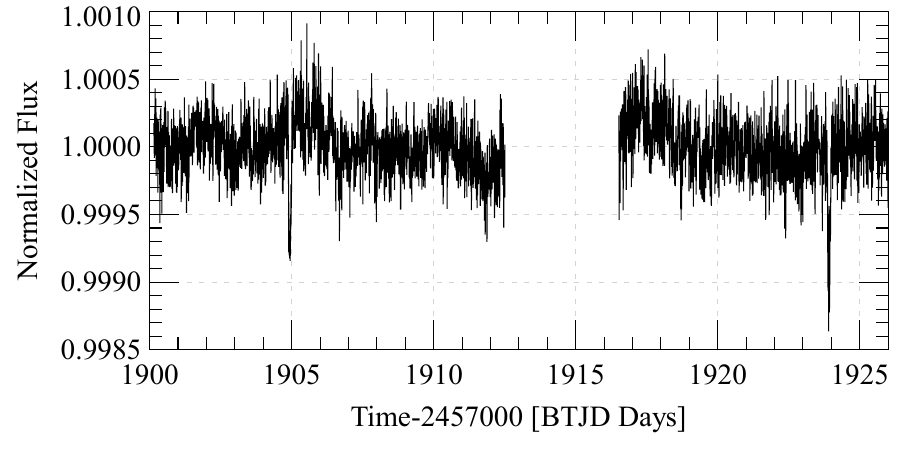}\caption{\label{fig:TESS} The TESS PSCSAP 2~min lightcurve binned into 10~minute bins. The known planet \thisplanet has transits at days 1905 and 1924, as well as a third transit during the gap.}
    	\end{center} 
\end{figure}

The \textit{TESS} mission \citep{ricker2015} observed \thisstar during Sector~22 for a total duration of approximately 23 days (the full Sector duration of $\sim 27$ days, minus $\sim 4$ day gap from d13 to d17). The normalized PDCSAP lightcurve is presented in Figure~\ref{fig:TESS}.  We first use the short cadence (2~minute) mission processed PDCSAP time series data to look for signs of rotation due to starspots rotating in and out of view on the surface of the star.  We find that the average brightness of HD\,97658 is stable with a RMS (root mean square) of 385~ppm, having no evidence of long-term variability due to spots during this observation period.  This conclusion is consistent with {\it TESS} observation period being shorter than the derived rotation period $P_{\rm rot} = 34 \pm 2$~d from the spectroscopic analysis of the Calcium H and K lines ($S_{\mathrm{HK}}$) \citep{guo2020}.

\begin{figure}
    	\begin{center}
    		\plotone{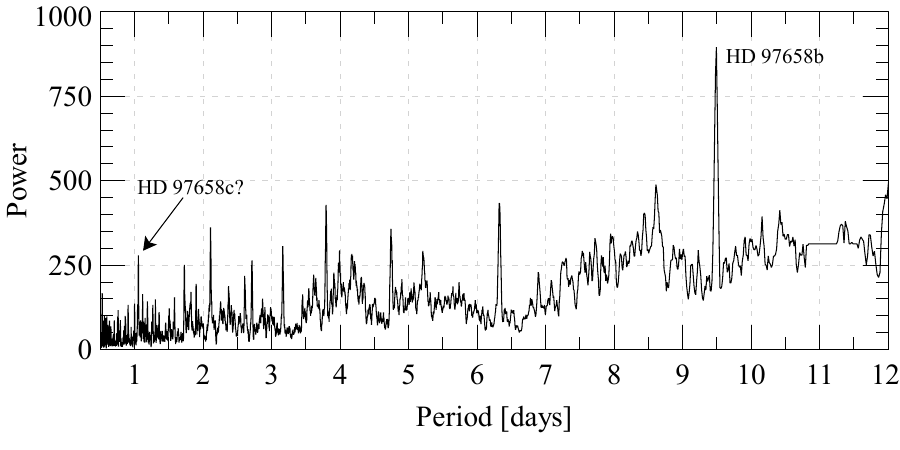}\caption{\label{fig:BLS} The Box Least-Squares periodogram of the TESS lightcurve of \thisstar. The strongest peak coincides with the known planet \thisplanet\ with a period of $\sim~9.5$~days. Most of the other peaks are harmonics of \thisplanet\!, but there is a notable spike at 1.05 days which is explored as a potential planet candidate. There is a peak at $\sim3.8$~days which proved insignificant upon further inspection. The other peaks are harmonics of the 9.5 and 1.05 day signals.}
    	\end{center} 
    \end{figure}

Next, we search for transits in the {\it TESS} data by computing a Box Least-Squares (BLS) periodogram \citep{Kovacs2002}, shown in Figure~\ref{fig:BLS}, using the \texttt{astropy} package \footnote{\url{https://docs.astropy.org/en/stable/timeseries/bls.html}}. We identify the known transiting planet HD\,97658~b, with an approximate period of 9.474~d. We note that only two transits of \thisplanet are detected in TESS data, while the third falls within the data gap.  We use \texttt{EXOFASTv2} \citep{eastman2017} to simultaneously fit the orbital parameters to the {\it TESS} time series along with the full radial velocity (RV) data series from \citet{dragomir2103}.  The program also simultaneously fit the MIST evolutionary models to estimate stellar properties \citep{CHOI2016MIST}.

\texttt{EXOFASTv2} reported a transit depth of \thisplanet\ of $712\pm38$~ppm with a duration of $2.80\pm0.04$~hours, centered at BJD $2458904.9366\pm0.0008$. With our measurement of the stellar radius and the transit depth, we compute a planet radius of $2.12\pm0.06$~\rearth. The resultant temperature of \thisplanet\ can then be found as $T_{\rm{eq}}=T_\star \sqrt{R_\star/2a}$, neglecting albedo. We find the  equilibrium temperature for planet b of 750$\pm$13~K. Our estimation of the equilibrium temperature is in line with the estimate from \citet{vangrootel2014}. 
%First first-author paper, woot! I'd like to thank the academy...
Using the planetary mass found with \texttt{EXOFASTv2}, 7.5$\pm0.9$~\mearth, and the planetary radius derived from our direct measurement of the stellar radius and the TESS transit depth, we determine a density of $\rho_\text{p}=3.7\pm0.5$~g~cm$^{-3}$. The density is consistent within 1~$\sigma$ of \citet{dragomir2103} and \citet{vangrootel2014}. This fitting of the RV and transit data yielded a semi-amplitude of $2.8\pm0.3$~m/s, which is also within 1-~$\sigma$ of \citet{dragomir2103}. The low amplitude RV signal largely limits the uncertainty in the measurements of planet b's density and any improvements are due to the decreased uncertainty in the transit depth and stellar radius. 

\begin{figure}[H]
    \centering
    \includegraphics[width=0.9\linewidth]{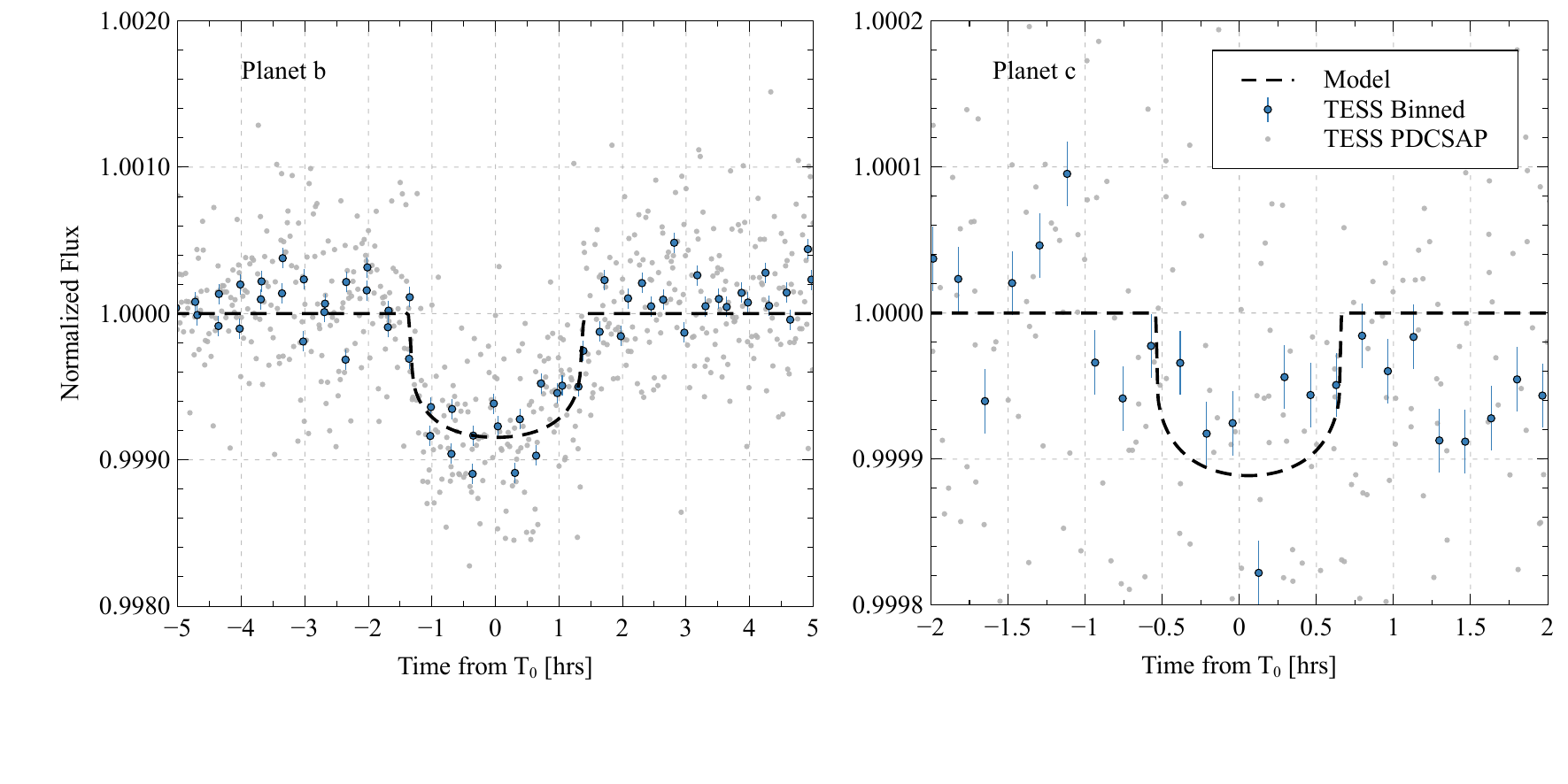}
    \caption{All of the TESS photometry is shown as transparent grey points. The folded and 10~minute binned TESS photometry with associated errors are shown as blue points. The transit models from \texttt{EXOFASTv2} are the dashed black line. \textbf{Left}: The best fit model for planet b with period 9.4896~days. \textbf{Right}: The \texttt{EXOFASTv2} transit model for the planet candidate ``c'' identified in Figure~\ref{fig:BLS} with period 1.054 days is shown as the blue dashed line. See \S~\ref{sec:exoplanet} for details. Note that the depth is approximately an order of magnitude less than the left plot.}
    \label{fig:folded}
\end{figure}

Finally, we use the {\it TESS} time series to look for any additional transiting planets.  Interestingly enough, the BLS periodogram analysis comes up with a signal at 1.054~d.  We use \texttt{EXOFASTv2} to model this candidate signal as an additional planet in the system at the same time as planet b. If such a planet candidate exists, \texttt{EXOFASTv2} indicates it would have a period of $1.05443179+_{-0.00000018}^{+0.00000011}$~days and cause a transit depth of $88\pm17$~ppm lasting 1.36~hours. The $T_0$ for the model was found as BJD $2458907.1\pm0.3$. Given this depth and our measurement of the stellar radius, the planet candidate would have a radius of 0.74~\rearth. The planet would be located at 0.019~au with an equilibrium temperature of 1565~K, found using the same method as for planet b. The transit model overlaps with the known planet transit and reduces the depth for planet b to $674\pm38$ and the corresponding radius to 2.06$\pm0.06$~\rearth. The folded lightcurve for this planet candidate is shown in Figure~\ref{fig:folded} right panel. The 10~minute binned out of transit photometry has a root mean square deviation of 39~ppm, about half of the transit depth.

In attempts to verify this signal with existing data, we searched the RV data for a signal with a period of 1.05~days , but did not find evidence to support it.  However, we note that a transiting planet with this  period and transit depth would not be massive enough to induce reflex motions detectable with current RV instruments/observations. We  compared the Bayesian Information Criterion (BIC), a measure of the goodness of the fit that penalizes over-parameterization, of the single planet model to the two planet model \citep{Schwarz1978-BIC}.  We find there is a small preference to the two planet model $\Delta BIC\simeq-8$. However, we still present this signal as only a candidate due to the low strength of the transit signal and lack of RV corroboration.

While beyond the scope of this paper, we propose that further investigation of this candidate would benefit from including archival time series over much longer time baselines than the TESS Sector 22 data presented here.  In particular, enabling transit timing variation analysis in the orbital fit for planet b in the model could provide additional independent evidence for the putative companion ``c".

\subsection{Star and Planet Properties from Transit Observables} \label{sec:propsfromobs}
\cite{Seager2003} demonstrated that combining the transit depth $\Delta F$, duration $\tau$ and period $P$ derived from the exoplanet transit lightcurve analysis yield the stellar density. Thus, with a direct determination of the stellar radius through interferometry, we can then directly obtain the stellar mass. This method has been applied to 55 Cnc \citep{Crida2018, Crida2018RNAAS} and HD 219134 \citep{Ligi2019}, for which the joint likelihood of the stellar mass and radius (\Ld$_{MR_\star}=\mathcal{L}_{MR\star}(\rho_{\star}, R_{\star})$) is expressed through the probability density function (PDF) of the density and radius \citep[see Eq. (2) of ][]{Ligi2019}. The PDF of the radius is itself expressed as a function of the PDF of the observables $\theta_{\rm LD}$ (angular diameter) and $d$ (distance), considered as Gaussian.

This yields the PDF of the planetary mass and radius, that depends on $\Delta F$, $P$, the semi-amplitude of the RV measurement $K$, and \Ld$_{MR_\star}$. Importantly, this method also allows computation of the correlation between parameters. This prevents, for example, determining absurd planetary densities that would not correspond to a realistic planetary mass.

Using this technique, and considering only planet $b$, we obtain $\rho_\star$=\rhostardir~g~cm\textsuperscript{-3}\, which yields $M_\star$~=~\mstardir~\msun, with a correlation of Corr($R_\star$,$M_\star$)=0.41. This low correlation is due to the high uncertainty on the stellar density. This direct determination of the mass is higher but consistent with those obtained with the different stellar  evolutionary models.

Applying the stellar mass derived from the transit model, the interferometrically derived stellar radius, transit model, and the RV semi-amplitude we find the planetary mass $M_\text{p}$=8.3$\pm$1.1~\mearth\ and radius $R_\text{p}$=2.12$\pm$0.06~\rearth, with the corresponding density $\rho_\text{p}$=4.8$\pm$0.7~g~cm$^{-3}$. These measurements are in good agreement with those found from the \texttt{EXOFASTv2} analysis. We note that the correlation between the planet's mass and radius is very low (Corr($R_\text{p}$, $M_\text{p}$) = 0.09). This is explained by Corr($R_\star$,$M_\star$), which is already low, and the high uncertainty on the transit and RV measurements parameters. Observations with higher precision are needed to reduce these uncertainties and increase the correlation between the parameters.

\begin{deluxetable}{lrl}\tablewidth{4in}
\tablecaption{Summarized Properties of the HD 97658 System\label{tab:props}}
\tablehead{\colhead{Property} & \colhead{Value} & \colhead{Source}} 
\startdata
\multicolumn{3}{c}{Measured Stellar Properties}\\\hline
Parallax [mas]                              &   46.412$\pm$0.022        & \citet{GAIAEDR3,Lindegren2021}\\
Distance [pc]                               &     $21.546\pm0.011$      & \citet{GAIAEDR3,Lindegren2021}\\
$[$Fe/H$]$ [dex]                            &   $-0.23\pm0.03$          & \citet{Howard2011}\\
$\theta_{\text{UD-R}}$ [mas]                &   \udd                    & \S \ref{sec:diam} Interferometry\\
$\theta_{\rm LD}$ [mas]                     &   \ldd                    & \S \ref{sec:diam} Interferometry\\
Linear Limb Darkening $\mu_\text{R}$        &   \mulimb                 & \S \ref{sec:diam} \citet{LDTK1, LDTK2}\\
$R_\star$ [R$_\sun$]                        &   \rstar                  & \S \ref{sec:diam} Interferometry, Parallax\\
$F_{\rm Bol}$ [erg s$^{-1}$ cm$^{-2}$]      &   \fbol                   &   \S \ref{sec:fbol_teff} SED Templates\\
\teff [K]                                   &   \temp                   & \S \ref{sec:fbol_teff}  Interferometry, SED\\
$L_\star$ [L$_\sun$]                        &   \lstar                  & \S \ref{sec:fbol_teff} $F_{\rm Bol}$, Parallax\\\hline
\multicolumn{3}{c}{Isochrone Properties --- \S~\ref{sec:ageandmass}}\\\hline
Age [Gyr]                                   & $3.9_{-2.03}^{+2.6}$      &  Combined Isochrone Models\\
M$_\star$ [\msun]\tablenotemark{a}          & $0.773_{-0.018}^{+0.015}$ & Combined Isochrone Models \\\hline
\multicolumn{3}{c}{\texttt{EXOFASTv2} Model Derived Properties --- \S~\ref{sec:exoplanet}}\\\hline
Transit Depth [ppm]                         & 712$\pm38$                &  \texttt{EXOFASTv2}\\
Period [days]                               &  $9.4897116\pm0.0000008$  & \texttt{EXOFASTv2}\\
$T_0$ [BJD]                                 &  $2458904.9366\pm0.0008$  & \texttt{EXOFASTv2}\\
$\text{R}_\text{p}/\text{R}_\star$          &  0.0267$\pm0.0007$        &  \texttt{EXOFASTv2}\\
Inclination [$\deg$]                        &  89.05$_{-0.24}^{+0.41}$  &   \texttt{EXOFASTv2}\\
Impact Parameter                            &  0.39$_{-0.18}^{+0.11}$   &  \texttt{EXOFASTv2}\\
Eccentricity                                &  0.05$_{-0.03}^{+0.04}$   &  \texttt{EXOFASTv2}\\
M$_\star$ [\msun]\tablenotemark{a}          & $0.75\pm0.02$             & \texttt{EXOFASTv2}, MIST \\
$M_\text{p}$ [\mearth]\tablenotemark{a}     & 7.5$\pm0.9$               & \texttt{EXOFASTv2}, MIST, \textit{K} \\
RV Semi-Amplitude $K$ [m/s]                 & $2.8\pm0.3$               & \texttt{EXOFASTv2}, RVs from \citep{dragomir2103} \\  
$a/R_\star$                                 & $24.2\pm0.7$              & \texttt{EXOFASTv2}\\
$R_\text{p}$ [R$_\earth$]                   & 2.12$\pm 0.06$            &  Transit Depth, Interferometric $R_\star$\\
$\rho_\text{p}$ [g cm$^{-3}$]               & $3.7\pm0.5$               & Transit Derived $R_\text{p}$, $M_\text{p}$\\
$T_{\rm{Eq}} [\text{K}]$                    & $751\pm12$                & \texttt{EXOFASTv2}, $a/R_\star$,  \teff \\\hline
\multicolumn{3}{c}{Stellar and Planetary Properties from Transit Observables --- \S~\ref{sec:propsfromobs}}\\\hline
$\rho_\star$ [g cm$^{-3}$]                   &   \rhostardir            & Transit Observed Properties\\
M$_\star$ [M$_\sun$]\tablenotemark{a}       &   \mstardir               & Interferometric $R_\star$, Transit Derived $\rho_\star$\\
$\log(g)$ [cgs]                             &   4.64$\pm$0.04           & Interferometric $R_\star$, Transit Derived $\rho_\star$\\
Corr($R_\star$, $M_\star$)                   &   0.41                   & \\\hline
$M_\text{p}$ [\mearth]\tablenotemark{a}     & 8.3$\pm1.1$               & Transit Derived $M_\star(\rho_\star, R_\star)$ and $K$\\
$\rho_\text{p}$ [g cm$^{-3}$]               & $4.8\pm0.7$               & Transit Derived $M_\text{p}$, $R_\text{p}$\\
Corr($R_\text{p}$, $M_\text{p}$)            &   0.09                    & \\
\enddata

\tablenotetext{a}{The table reflects the computed mass of the star and planet with two different methods. See \S~\ref{sec:derived} for more details.}
\end{deluxetable}

%%%%%%%%%%%%%%%%%%%%%%%%%%%%%%%%%%%%%%%%%%%%%%%%%%%%%%%%%%%%%%%%%%%%%%%%%%%%%%%%
\section{Conclusion} \label{sec:conclusion}
%%%%%%%%%%%%%%%%%%%%%%%%%%%%%%%%%%%%%%%%%%%%%%%%%%%%%%%%%%%%%%%%%%%%%%%%%%%%%%%%
In this work we use the GSU CHARA interferometric array to obtain a highly precise measurement of the angular diameter of \thisstar. We combine measurements from both the infrared Classic instrument as well as the optical PAVO and VEGA instruments for more complete coverage of the UV plane, which helps ensure a well defined angular diameter. We also combine photometric measurements from a panoply of sources to find bolometric flux with SED fitting. These two measurements allow an estimation of temperature independent of distance measurements.

We provide the most direct measurement of the star's radius which paired with the Gaia parallax produce a $\sim$1\% uncertainty in the physical radius and a $\sim$~0.5\% uncertainty in effective temperature. Previous works exploring the properties of \thisstar were able to obtain estimates of the stellar radius and temperatures which are in good agreement with the measurements performed in this work. Because of this we cannot report substantially different composition and properties of HD 97658 b, but we can provide greater certainty in the previous results.

Follow-up observations of \thisstar with  JWST will allow a more precise and accurate measurement of the transit depth.  This is a particularly interesting measurement to pin down as current best measurements of the transit depth are accurate to only $\sim 5$\%, which complicates more accurate analysis of the planet. These follow up observations would provide further exciting insight into this nearby super-Earth planet. We also eagerly await the Magdalena Ridge Observatory interferometer which will  enable observations of fainter targets and baseline bootstrapping which will ease optimal sampling of the UV plane \citep{MROI}.

%%%%%%%%%%%%%%%%%%%%%%%%%%%%%%%%%%%%%%%%%%%%%%%%%%%%%%%%%%%%%%%%%%%%%%%%%%%%%%%%
\section*{Acknowledgements}
\acknowledgments

We offer our sincere appreciation to the observing team, scientists, and support staff at the CHARA Array. This work is based upon observations obtained with the Georgia State University Center for High Angular Resolution Astronomy Array at Mount Wilson Observatory. The CHARA Array is supported by the National Science Foundation under Grant No. AST-1636624 and AST-1715788.  Institutional support has been provided from the GSU College of Arts and Sciences and the GSU Office of the Vice President for Research and Economic Development. The VEGA time at the CHARA Array was granted through the NOIR Lab community access program (PI: Ligi; 2017A-0162, 2018A-0178, 2018B-0019, 2020A-0172).

R.L. has received funding from the European Union's Horizon 2020 research and innovation program under the Marie Sk\l odowska-Curie grant agreement n. 664931 and from the European Research Council (ERC) under the European Union’s
Horizon 2020 research and innovation program (grant agreement CoG \#683029)

We thank Dan Huber and Tim White for their time so willingly given to make this project better. We thank Pierre Maxted for their assistance in compiling and running \texttt{bagemass}. We thank Jason Eastman for their assistance with \texttt{EXOFASTv2}. Finally, we offer our appreciation to the anonymous reviewer whose thorough comments helped make this article its best possible realization.

This work uses results from the TESS mission. Funding for the TESS mission is provided by NASA's Science Mission directorate.

This research made use of the Jean-Marie Mariotti Center JSDC catalogue\footnote{available at \url{http://www.jmmc.fr/catalogue\_jsdc.htm}}.

This research made use of the SIMBAD literature database, operated at CDS, Strasbourg,France, and of NASA’s Astrophysics Data System.

This work presents results from the European Space Agency (ESA) space mission Gaia. Gaia data are being processed by the Gaia Data Processing and Analysis Consortium (DPAC). Funding for the DPAC is provided by national institutions, in particular the institutions participating in the Gaia MultiLateral Agreement (MLA). The Gaia mission website is \url{https://www.cosmos.esa.int/gaia}. The Gaia archive website is \url{https://archives.esac.esa.int/gaia.}

This publication makes use of data products from the Two Micron All Sky Survey, which is a joint project of the University of Massachusetts and the Infrared Processing and Analysis Center/California Institute of Technology, funded by the National Aeronautics and Space Administration and the National Science Foundation.

This publication makes use of data products from the Wide-field Infrared Survey Explorer, which is a joint project of the University of California, Los Angeles, and the Jet Propulsion Laboratory/California Institute of Technology, funded by the National Aeronautics and Space Administration.

This research made use of Lightkurve, a Python package for Kepler and TESS data analysis (Lightkurve Collaboration, 2018).

\vspace{5mm}
\facilities{CHARA (PAVO, VEGA, Classic)}

\software{Scipy \citep{Jones2001}, astropy \citep{astropy2018}, Redfluor \& Calibir, PAVO \citep{Ireland2008}, vegadrs \citep{Mourard2009, Mourard2011},  bagemass \citep{Maxted2015}, EXOFAST v2 \citep{eastman2017}. Limb Darkening Toolkit \citep{LDTK1, LDTK2}, Lightkurve (and dependencies) \citep{lightkurve}, celerite\citep{celerite}, Astrocut \citep{tesscut}, Astroquery \citep{astroquery} }
\section*{Data Availability}
The data underlying this article will be shared on reasonable request to the corresponding author. All interferometric data are available in the CHARA archive \footnote{\url{http://www.chara.gsu.edu/observers/data-policy-access}}.
\bibliography{bib}

\end{document}